\documentclass[12pt,a4paper]{article}
\usepackage[utf8]{inputenc}
\usepackage{amsmath}
\usepackage{amsfonts}
\usepackage{amssymb}
\usepackage{graphicx}

\usepackage{multirow}

\providecommand{\keywords}[1]{\textbf{Keywords---} #1}

\usepackage[labelformat=simple]{subcaption}

\usepackage{url}

\usepackage[nomarkers,nolists,figuresfirst]{endfloat}

\usepackage{authblk}
\usepackage{color}
\usepackage{epstopdf}
\usepackage{wrapfig}

\usepackage{pgfplots}
\usepackage{tikz}
\usetikzlibrary{shapes}
\usetikzlibrary{arrows}

\usepackage[
square,
sort,
comma,
numbers
]{natbib}

\title{\vspace{-4cm} High fidelity blood flow in a patient-specific arteriovenous fistula}
\author[1]{McCullough, J.W.S.}
\author[1,2]{Coveney, P.V.$^*$}

\affil[1]{Centre for Computational Science, Department of Chemistry, University College London, UK}
\affil[2]{Informatics Institute, University of Amsterdam, Netherlands}

\date{\today}

\begin{document}
\maketitle

\begin{abstract}
An arteriovenous fistula, created by artificially connecting segments of a patient's vasculature, is the preferred way to gain access to the bloodstream for kidney dialysis. The increasing power and availability of supercomputing infrastructure means that it is becoming more realistic to use simulations to help identify the best type and location of a fistula for a specific patient. We describe a 3D fistula model that uses the lattice Boltzmann method to simultaneously resolve blood flow in patient-specific arteries and veins. The simulations conducted here, comprising vasculatures of the whole forearm, demonstrate qualified validation against clinical data. Ongoing research to further encompass complex biophysics on realistic time scales will permit the use of human-scale physiological models for basic and clinical medicine.

\end{abstract}


\section{Introduction}
\label{sec:Intro}
The development of the virtual human to recreate an individual's specific physiological processes in a digital format is a major goal of computational biomedicine, heralding truly personalised medicine along with `healthcasts' \cite{Hunter2013,Muszkiewicz2016,Biglino2017,Hoekstra2018,Kim2019,Morton2019,Hokken2020}.  The increasing capability and power of computational software and hardware has allowed significant steps to be made towards this goal \cite{Hunter2010, Hunter2013}. The development of such a `digital doppelganger' is not a strict progression from low-level (e.g. genomic and DNA studies) to high-level (e.g.\ organ modelling or epidemiological phenomena) processes but both top-down and bottom-up \cite{Noble2012}. The purpose of the present paper is to report a further advance to these efforts, specifically in the field of blood flow modelling. While 3D flow in large arterial structures has been simulated \cite{Xiao2013,Randles2015}, we demonstrate how scalable 3D modelling codes can be used to examine macroscopic blood flow in coupled arteries and veins at human scale. We show how the capillary beds linking the two networks in a virtual human can be captured in an efficient manner and applied to human-scale flow problems.\\ 

Blood flow is a complex and multiscale process that is fundamental to the operation of the body's physiological processes. From a simulation perspective however, the macroscopic nature of blood may be treated as a continuum fluid that behaves according to the Navier-Stokes equations for fluid flow. Many studies solve these equations for vascular networks through the use of highly simplified 1D representations. The specific properties of an individual's vascular network are used as input for nodal locations within such 1D models \cite{Sheng1995,Qureshi2014,Muller2014,Mynard2015b}. This approach has advantages relating to the relative ease in parameterisation of the model and the computational effort required to reach a solution. In comparison, 3D models permit a far higher fidelity study of the blood flow by means of an exact representation of the domain of interest and do not rely on major assumptions about flow behaviour (e.g. symmetry) that are made by 1D solvers \cite{Xiao2013,Randles2015,mccullough2020blood}. However, 3D simulations are more demanding to execute. As supercomputers become more powerful, it is more tractable to study 3D flow of large, human-scale, vascular domains \cite{Xiao2013,Randles2015,mccullough2020blood}. Studies of this nature are better equipped to reveal the subtle, patient-specific, phenomena that exist within complex vasculatures. To take advantage of increasing computational performance, a numerical method that demonstrates  excellent parallel scaling capability whilst studying complex geometries is essential. In the development of the virtual human, such performance will allow an \textit{in silico} model to generate necessary data and provide this to a patient's healthcare team in a timely fashion. The lattice Boltzmann method (LBM) used in this work is an ideal candidate for meeting these requirements due to its easily parallelised algorithm and often localised approach to solving the complex geometries or moving boundaries found at multiple scales in blood flow \cite{Groen2013,Groen2018,Latt2020}.\\ 

Irrespective of whether a 1D or 3D model is used to represent the systemic blood vessels, there will remain sections of the vascular network, vital for physiological function, that must be described using sub-grid models. A typical instance of this is the flow through the capillary beds that feed the muscles and skin throughout the body. These vessels are too small to be resolved with typical scanning techniques used to obtain geometries of larger vessels. For instance, typical MRI scans used in hospitals can only resolve features larger than 1 mm \cite{Cleary2014}. Alternate scanning techniques such as computed tomography (0.5 mm) and ultrasound (0.15 mm) may achieve improved resolution though at the cost of lower contrast resolution \cite{Lin2009}. The flow behaviour within these unresolved vessels becomes incorporated into the boundary conditions of macroscopic flow models. The development of a self-coupled version of the LBM code used in this study, HemeLB, has been used to show how this framework can simulate flow in human-scale arteries and veins \cite{mccullough2020blood}. Additionally, we provide further details on the modelling methods used in this current work in Section \ref{sec:Methods}.\\

\begin{figure}[!ht]
   \centering
   \includegraphics[width=0.98\textwidth]{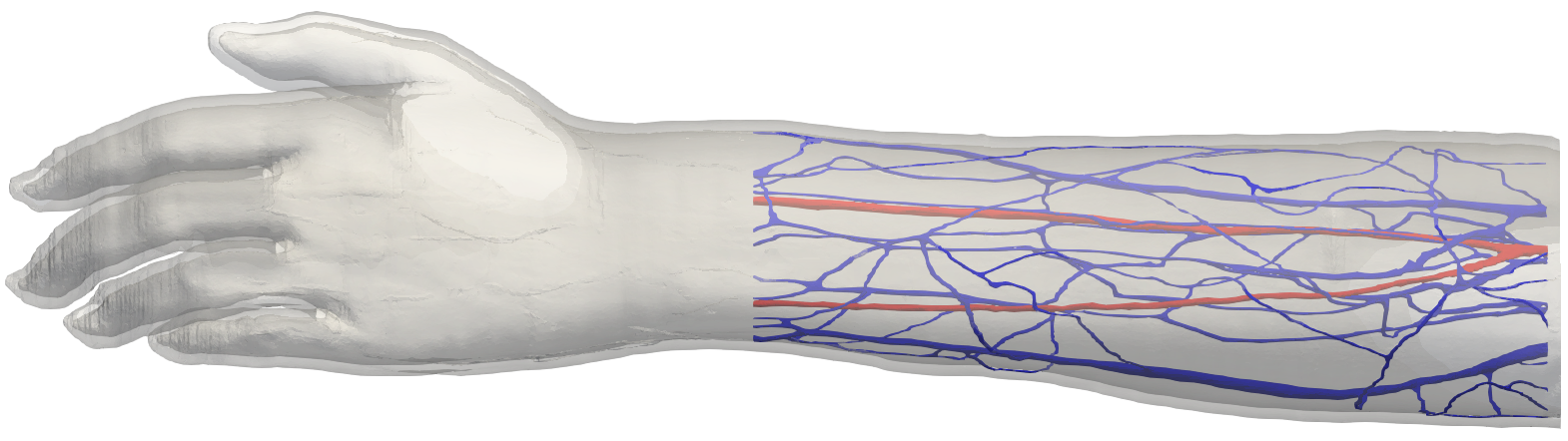} 
  \caption{Human scale vessels in the left forearm. Arteries are coloured red and the veins in blue. The technique used to capture the 3D vessels was able to resolve vessels to a diameter of approximately 0.3 mm. The vessels were truncated at the wrist as shown, the vessels in the hand were represented within the sub-scale material model connecting the simulations of the arteries and veins.}
  \label{LeftForeArm}
\end{figure}

We provide a contextual setting for this work by examining 3D flow through a personalised representation of the major arteries and veins in the human forearm. For patients suffering from kidney failure who require haemodialysis to remove waste products from the bloodstream, an arteriovenous fistula is a common course of action to facilitate treatment \cite{Allon2002,NHSdial2019}. This procedure often links the cephalic vein in the forearm to the radial artery, although other connections may also be selected. Bypassing the peripheral vasculature increases the blood flow through the vein, in turn increasing the size and strength of the vessel. Once matured, the vein is then used as an access point for blood to be removed for dialysis and returned to the body. The utilisation of \textit{in silico} models prior to treatment permits surgeons to assess different treatment options prior to undertaking invasive procedures \cite{Allon2002,NHSdial2019}. In this work we examine flow through the left forearm vessels illustrated in Figure \ref{LeftForeArm}. The technique used to capture the 3D vessels was able to resolve vessels to a diameter of approximately 0.3 mm. Vascular structures smaller than this, such as the palmar arch, are represented within the sub-scale material model connecting the simulations of the arteries and veins. In our study, we firstly study flow in unmodified vessels to assess baseline parameters prior to treatment. Following this we demonstrate how the model can be used to assess a candidate fistula location and its impact on vessels throughout the arm.\\

Within the literature, there exists a number of \textit{in silico} models for studying arteriovenous fistula creation and its application to surgical planning and evaluation, for example \cite{Bode2012,Decorato2014,Manini2014,McGah2014,Zonnebeld2017,Bai2019}. Of these, cases that capture large-scale vascular domains are typically 1D studies \cite{Bode2012,Manini2014,Zonnebeld2017} whilst 3D studies \cite{Decorato2014,McGah2014, Bai2019} typically only focus on the region of the fistula and do not consider the wider arterial and venous networks throughout the arm. In our model we are able to combine these two features by efficiently simulating 3D resolved flow in detailed networks of large-scale vessels. This is an important step as factors dictating the success of a fistula depend on flow in both the created junction and throughout the remainder of the arm.  \\

In what follows we discuss the formulation of our \textit{in silico} model. We use this to analyse flow in normal arteries and veins within the left forearm and those that have been modified to incorporate an arteriovenous fistula. We compare our results to those reported in clinical studies and demonstrate excellent agreement.\\

\section{Results}
We applied the sub-scale method discussed in Section \ref{sec:SubScale} to a large model of human blood vessels. This domain consists of the left forearm of the full human Yoon-sun model \cite{ITIS2019} (Figure \ref{LeftForeArm}), which is available as part of the Virtual Population (ViP) library \cite{Christ2010, Gosselin2014}. The segmentation of the full human model domain meant that the finest vessels in this domain had a diameter of approximately 0.4 mm. We discretised these selected geometries to a resolution of $\Delta x$ = 50 $\mu m$, resulting in a total of 6,128,855 lattice sites in the arterial geometry and 16,767,772 in the venous geometry. We conducted simulations with a time step of $\Delta t$ = 3 $\mu s$, resulting in a LBM relaxation time of $\tau$ = 0.5144. This was utilised in a two relaxation time collision kernel with a `magic' parameter of $\Lambda$ = 1/12 \cite{Kruger2017}. These discretisation parameters were selected based on two key considerations. The first, governing $\Delta x$, was that the finest vessels were sufficiently well resolved to capture the flow within them. The second was that $\Delta t$ (and by corollary $\tau$) permitted the LBM method to stably solve flow within the expected physiological values. The values chosen meet our expectations for these criteria. All test cases were run for a total of 1.1 million time steps. This corresponds to 3.3 $s$ of physical time being simulated and demonstrates the capacity of our model to efficiently represent multiple cardiac cycles in addition to a period of numerical `warm-up' to prepare flow in the vessels. These were conducted on SuperMUC-NG and each took approximately 2 hours of wall time to run on 50 nodes (2400 cores). In further work, we also examined a similar vascular set discretised to $\Delta x$ = 25 $\mu m$, these contained 64,998,754 lattice sites in the arterial geometry and 166,652,137 in the venous geometry and took approximately 2.5 hours to run on 1500 SuperMUC-NG nodes. \\

In the long term study of patients with arteriovenous fistula, a key observational parameter is how the vessels change shape in response to its insertion. Once this vascular remodelling has stabilised and blood flow is sufficient for dialysis to occur, the fistula is said to have matured. As our model is currently not able to capture this process dynamically, we look to examine the changes to the vascular flow patterns caused by the instantaneous creation of the fistula. These changes may serve as indicators for the success of maturation. A future study conducting such simulations on a population of patients would help to understand this further.

To demonstrate that our model captures the flow behaviour expected in both coupled arteries and veins as well as within an arteriovenous fistula we consider the following cases. The first is a `healthy' case where we apply a typical flow waveform to the unmodified domains with a single capillary bed region (all venous inlets in the hand region are fed by the two arterial outlets). In the remaining cases we demonstrate how our model can capture an arteriovenous fistula. A common option for creating a fistula is to connect between the radial artery and cephalic vein \cite{VanTricht2005}. This connection bypasses the capillaries in the hand and serves to increase blood flow in the cephalic vein. In the first of our examples, we modify the arterial and venous geometries to mimic a surgically created end-to-side fistula \cite{VanTricht2005} both before and after maturation. We explicitly couple these modifications together within the coupling map. \\

Inlet flow to the arterial geometry in all cases was provided according to the profile in Figure \ref{InletVelocity}. This was generated based on data presented in \cite{Mynard2015b} for the right brachial artery. \\

\subsection{Healthy vessels}
To represent healthy vessels, we leave the entire geometry unmodified. The capillary bed consists of a single bed constructed by both arterial outlets and the 12 venous inlets in the hand region of the vascular structure. \\

\begin{figure}[!ht]
 \begin{subfigure}{0.49\textwidth}
   \centering
   \includegraphics[width=0.98\textwidth]{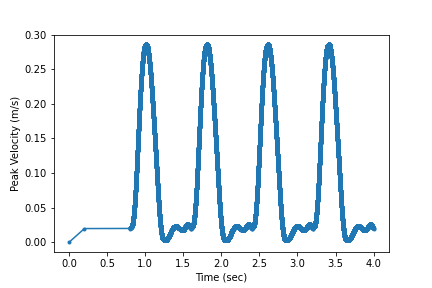} 
  \caption{Arterial inlet profile}
  \label{InletVelocity}
 \end{subfigure}
  \begin{subfigure}{0.49\textwidth}
   \centering
   \includegraphics[width=0.98\textwidth]{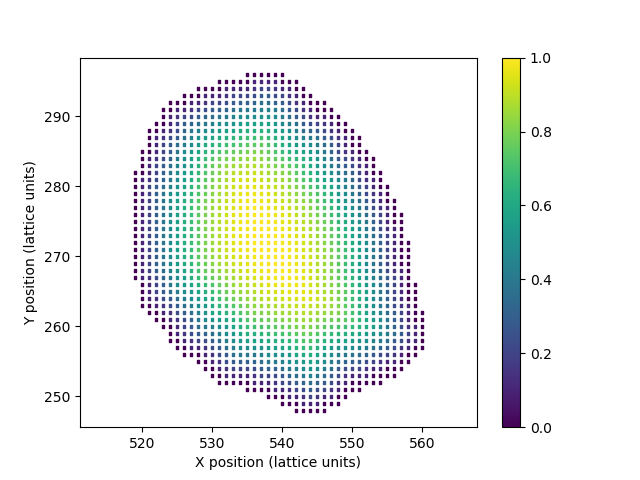} 
  \caption{Arterial inlet distribution}
  \label{InletDistribution}
 \end{subfigure}
 \begin{subfigure}{0.98\textwidth}
   \centering
   \includegraphics[width=0.98\textwidth]{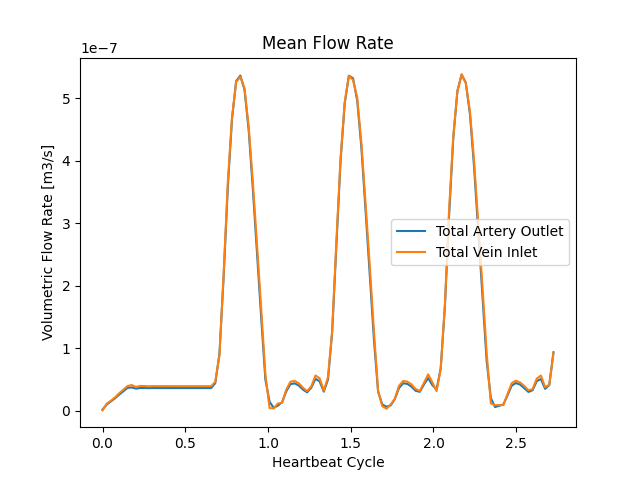}
   \caption{Total flow}
   \label{HealthyPCBinout}
 \end{subfigure}
  \caption{Simulation inlet conditions used in all studies and the observed response in the healthy vessels. \subref{InletVelocity} details the flow velocity provided to the inlet of the arterial geometry for the forearm flow cases. The first 0.8s of the flow represents an initial warm-up period of flow within the system.\subref{InletDistribution} indicates the distribution of scaling weights applied to the flow velocity at the inlet of the arterial geometry. \subref{HealthyPCBinout} compares the total inlet and outlet flow rate through the healthy vessels.}
  \label{HealthyInOut}
\end{figure}

In Figure \ref{HealthyInOut} we present the total flow rate entering and exiting the forearm domain being modelled. For the parameters demonstrated here, it can be observed that a steady-state flow response is obtained following the initial warm-up period.  \\

\begin{figure}[!ht]
\begin{subfigure}{0.98\textwidth}
   \centering
   \includegraphics[width=0.9\textwidth]{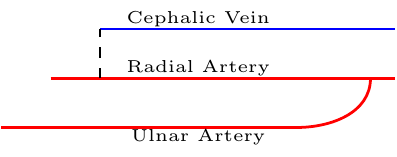}
   \caption{Schematic}
   \label{AVFschematic}
   \end{subfigure}
   \begin{subfigure}{0.98\textwidth}
   \centering
   \includegraphics[width=0.9\textwidth]{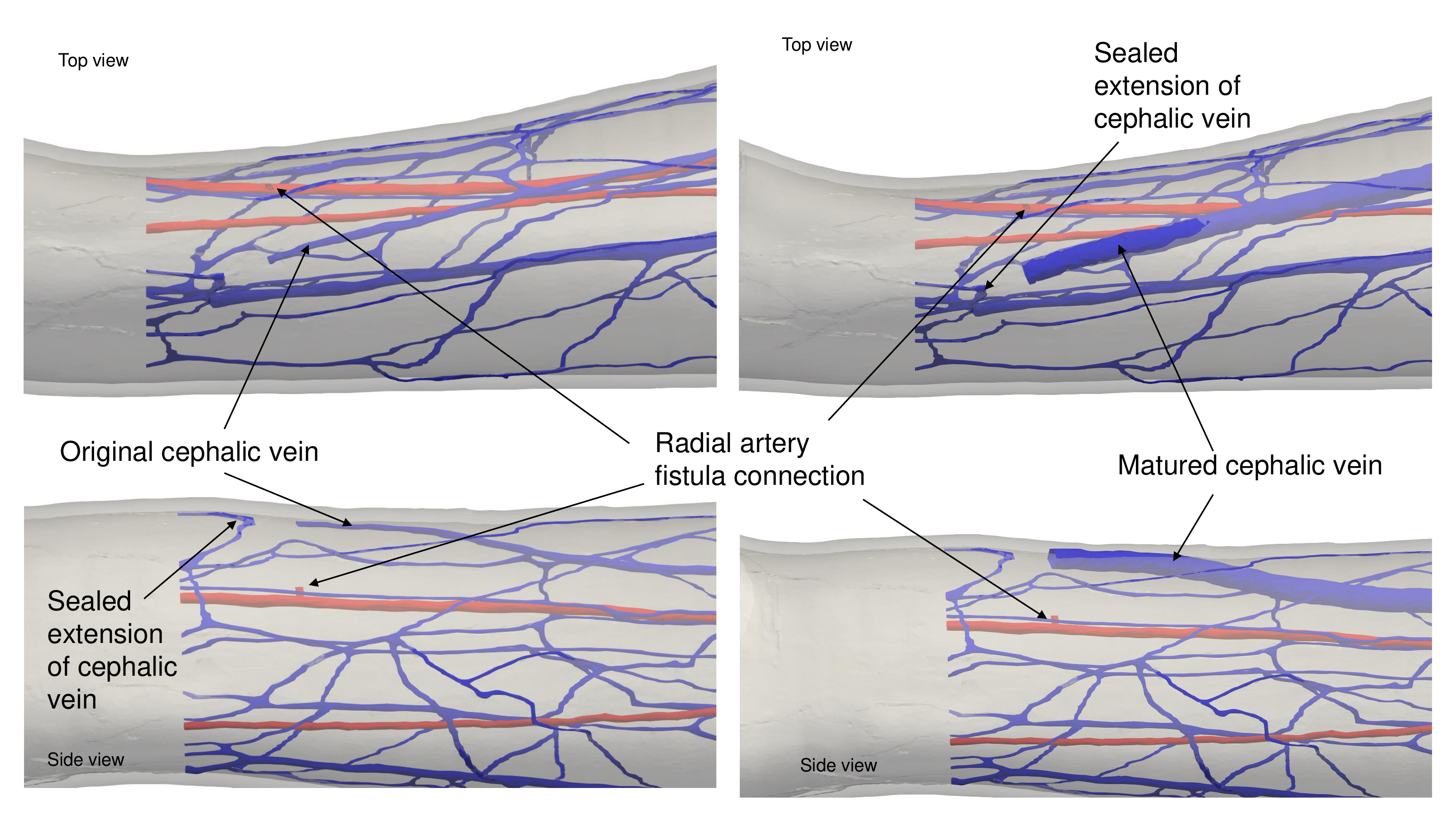}
   \caption{Vascular geometries}
   \label{AVFvessels}
   \end{subfigure}
  \caption{An arteriovenous fistula is created by artificially connecting sections of the arterial and venous vasculatures to alter the distribution of blood flow through the vessels. In \subref{AVFschematic} we provide a schematic map of the geometric modifications made to represent an end-to-side fistula. In this configuration, the cephalic vein is terminated early and attached to the radial artery. In \subref{AVFvessels} the realisation of these features in pre- and post-maturation geometries is highlighted.}
  \label{AVFlayout}
\end{figure}

In the following sections we explicitly modify the vascular geometry to incorporate an arteriovenous fistula, firstly immediately after surgical connection and then following a period of maturation. We consider the end-to-side option of joining the cephalic vein to the radial artery. These are illustrated in Figure \ref{AVFlayout} schematically alongside their realisation in the modified vascular domains.

\subsection{End-to-side arteriovenous fistula - pre-maturation}
In this setting, we modify the venous geometry such that the cephalic vein is terminated ahead of a junction and mapped in a 1:1 fashion to an opening made in the radial artery. The cephalic vein is kept at the same diameter as the healthy vessels to capture flow pre-maturation.\\

In Figure \ref{AVFFlowResults}, the total flow through the coupled domain again closely matches that provided at the input, indicating the conservation behaviour of the coupling scheme. The particular differences caused by the creation of the fistula can be seen more clearly in Figure \ref{AVFpreCephChange} which highlights the transient outflow of the major veins of the domain. Peak flow exiting via the basilic vein has roughly halved whilst the flow exiting the cephalic vein has increased by approximately two-thirds. Under this configuration, there is now almost equal flow exiting the two major vessels. \\

\subsection{End-to-side arteriovenous fistula - post-maturation}
In this setting, we modify the venous geometry such that the cephalic vein has dilated as a result of maturation. The mapping remains the same as before maturation with the transfer coefficients modified to reflect the change in size of the cephalic vein. In this dilated configuration, the diameter of the cephalic vein has increased by a factor of approximately 2.7. \\

\begin{figure}[!ht]
 \begin{subfigure}{0.49\textwidth}
   \centering
   \includegraphics[width=0.98\textwidth]{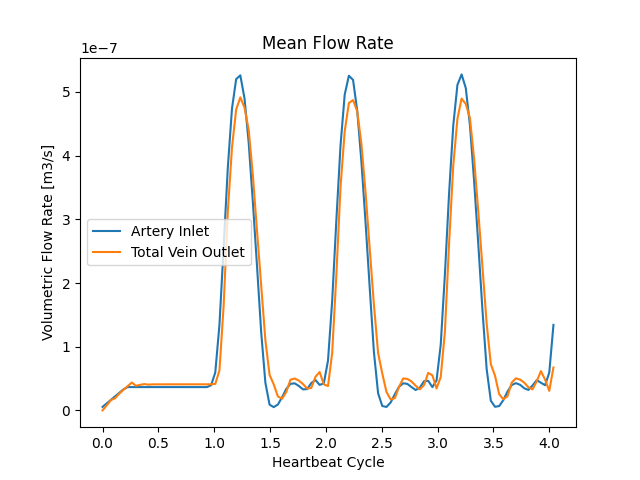}
   \caption{Total flow}
   \label{AVFpreinout}
 \end{subfigure} 
 \begin{subfigure}{0.49\textwidth}
   \centering
   \includegraphics[width=0.98\textwidth]{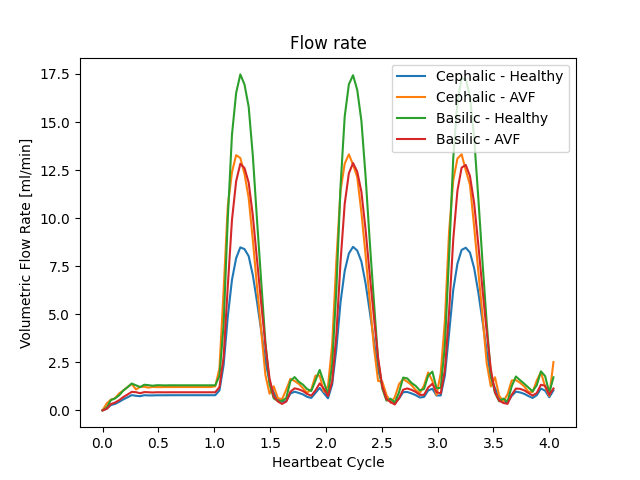}
   \caption{Flow in major vessels}
   \label{AVFpreCephChange}
 \end{subfigure}
 \begin{subfigure}{0.49\textwidth}
   \centering
   \includegraphics[width=0.98\textwidth]{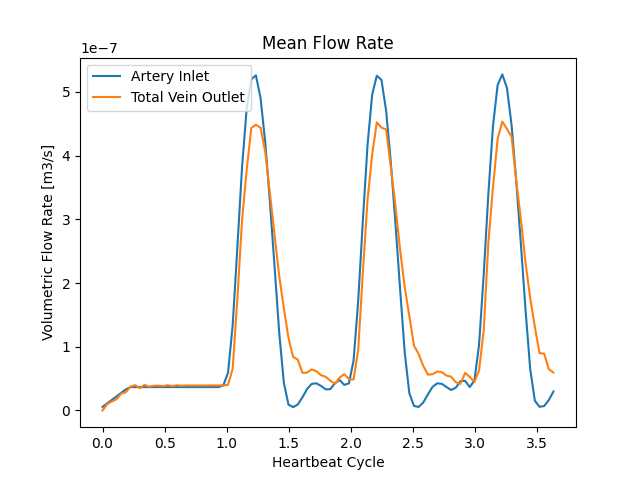}
   \caption{Total flow}
   \label{AVFpostinout}
 \end{subfigure} 
 \begin{subfigure}{0.49\textwidth}
   \centering
   \includegraphics[width=0.98\textwidth]{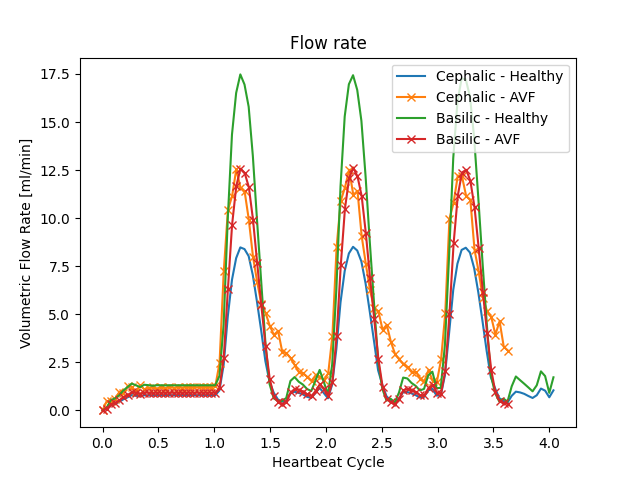}
   \caption{Flow in major vessels}
   \label{AVFpostCephChange}
 \end{subfigure}
  \caption{Flow rates through the forearm blood vessels with an arteriovenous fistula in its pre- and post-maturation states. Figures \subref{AVFpreinout} and \subref{AVFpostinout} indicate the total inlet and outlet flow of the system. Figures \subref{AVFpreCephChange} and \subref{AVFpostCephChange} compare flow exiting the major veins to that in the `healthy' vessels.}
  \label{AVFFlowResults}
\end{figure}

In Figure \ref{AVFFlowResults}, the change in similarities and differences in the flow patterns exiting the venous geometry can immediately be seen. During the periods of peak flow, the redistribution of the total flow compared to the healthy case is quite similar to that seen in the pre-maturation case. During the period of reduced driving flow, in the second half of the heartbeat profile, the flow exiting the cephalic vein reduces much more slowly then that seen in other cases. As the basilic vein profile remains similar to that seen in the pre-maturation case, this indicates that the larger cephalic vein is being supplemented by flow volume that would have exited the non-major veins.\\ 

\begin{figure}[!ht]
\centering %
 \begin{subfigure}{0.45\textwidth}
   \includegraphics[width=0.98\textwidth]{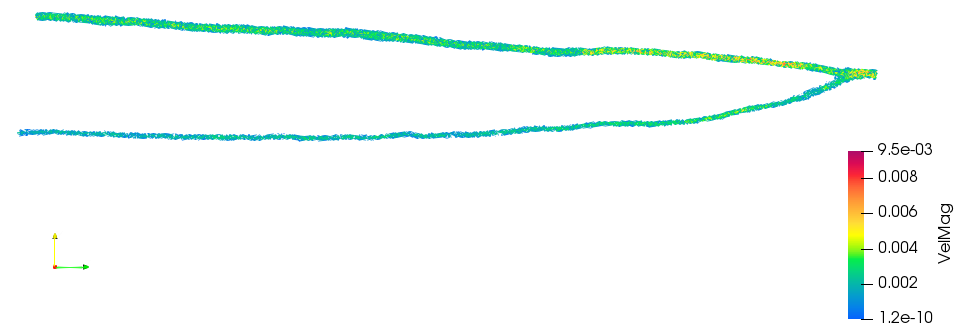}
  \caption{Arterial flow with healthy vessels} 
  \label{HealthyFlowA}
 \end{subfigure} 
 \begin{subfigure}{0.45\textwidth}
   \includegraphics[width=0.98\textwidth]{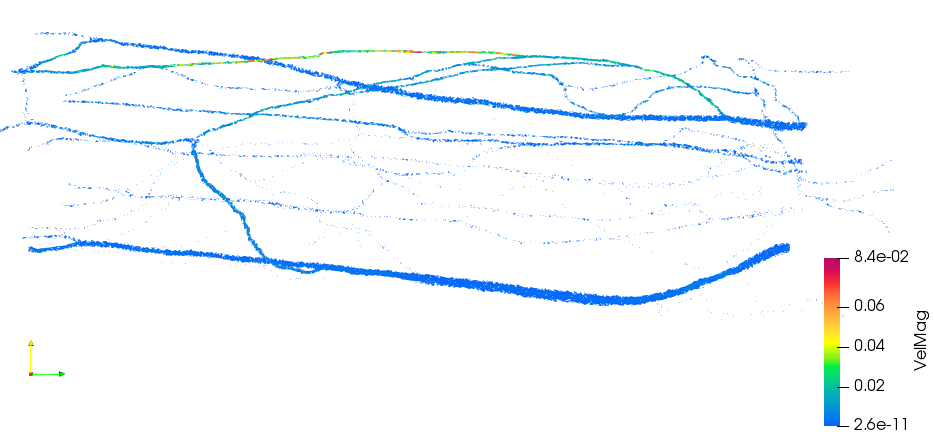}
  \caption{Venous flow with healthy vessels}
  \label{HealthyFlowV}
 \end{subfigure} 
  \begin{subfigure}{0.45\textwidth}
   \includegraphics[width=0.98\textwidth]{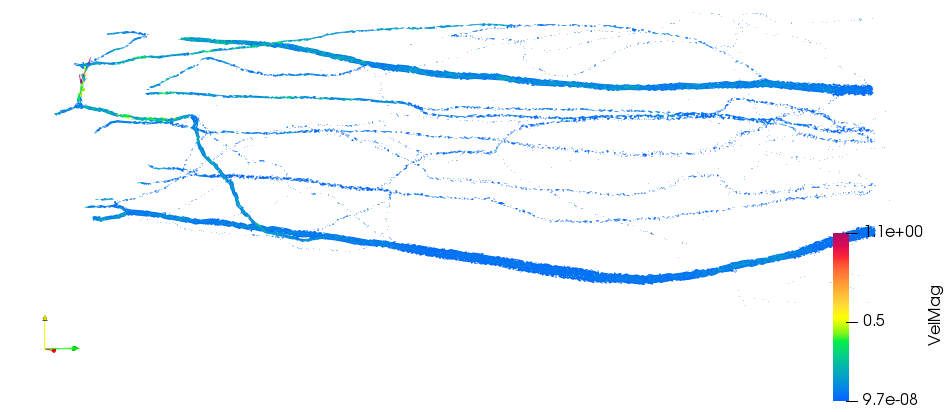}
  \caption{Venous flow with pre-maturation fistula}
  \label{AVFVeins}
 \end{subfigure}
  \begin{subfigure}{0.45\textwidth}
   \includegraphics[width=0.98\textwidth]{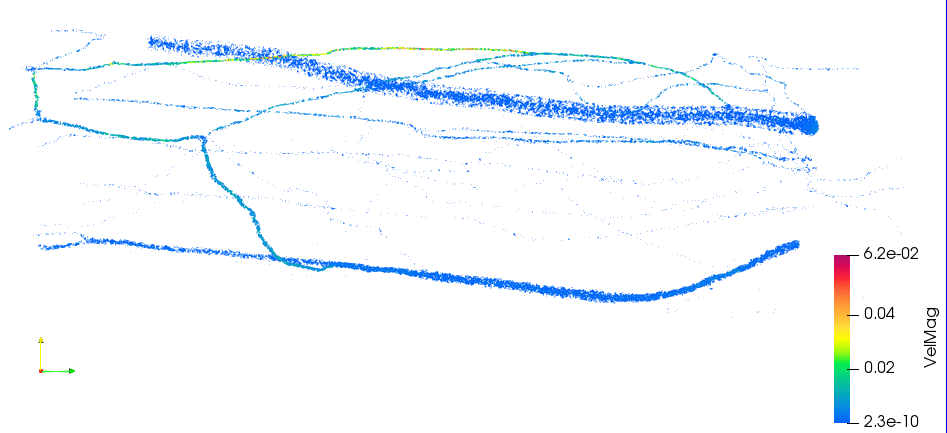}
  \caption{Venous flow with post-maturation fistula}
  \label{AVFdilateVeins}
 \end{subfigure}  
   \begin{subfigure}{0.45\textwidth}
   \includegraphics[width=0.98\textwidth]{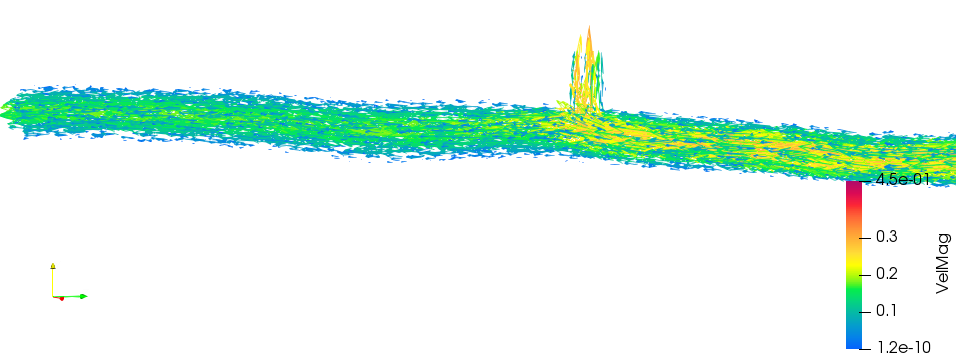}
  \caption{Arterial entry to fistula}
  \label{ArteryExitFlow}
 \end{subfigure} 
   \begin{subfigure}{0.45\textwidth}
   \includegraphics[width=0.98\textwidth]{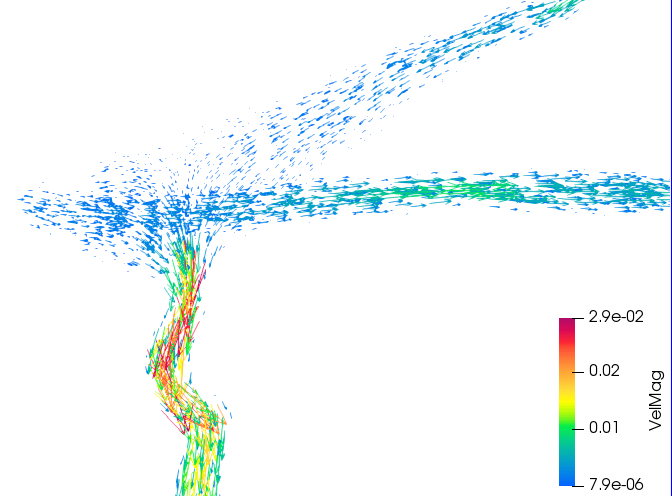}
  \caption{Venous flow in post-maturation fistula just below the sealed extension of the cephalic vein}
  \label{AVFdilateVeinsJunction}
 \end{subfigure}
  \caption{Snapshots of 3D flow after 700,000 time steps of simulation. In all cases the flow is represented by arrows oriented by velocity and coloured by velocity magnitude in m/s.  In \subref{HealthyFlowA} and \subref{HealthyFlowV}, we present flow in the healthy vessels where the preference for arterial flow in the radial artery is clearly evident due to the orientation of the vascular bifurcation. Figures \subref{AVFVeins} and \subref{AVFdilateVeins} represent the flow in the pre- and post-maturation veins respectively. In the venous cases, the highest flow velocities do not occur in the cephalic or basilic veins but in particularly narrow peripheral vessels. The alteration of flow due to the sealed extension of the cephalic vein (noted in Figure \ref{AVFvessels}) is obvious, particularly in the pre-maturation case. In a clinical setting, such simulation studies would allow such features to be noted and corrected ahead of any surgery being undertaken. In Figures \subref{ArteryExitFlow} and \subref{AVFdilateVeinsJunction} we provide zoomed images of flow features of particular interest. A key advantage of 3D simulation is to be able to generate and examine high fidelity representations of the simulated blood flow. For example, note the drop in flow velocity after the arterial exit in \subref{ArteryExitFlow}. }
  \label{FlowMaps}
\end{figure}

The use of 3D flow simulations permits far more detailed visualisation of flow than is possible in 1D studies. This can be particularly informative for flow in patient-specific geometries (as was highlighted in Section \ref{sec:Intro}). We provide examples of flow through the arteries and veins of the healthy vessels and those with geometry-modified arteriovenous fistula in Figure \ref{FlowMaps}. \\

\section{Discussion}
A particular challenge for the verification and validation of simulated 3D blood flow is the significant variation in the geometric and flow characteristics among humans. In the vessels examined in our study, the inlet diameter of the brachial artery providing flow to the full system was approximately 2.2 mm. In the healthy and pre-maturation configurations, the approximate diameter of the cephalic vein at the entry point of the fistula was 1.4 mm and at the exit plane 2.9 mm. In the matured vessels these measurements were 4.8 and 5.9 mm respectively. When compared to observational literature studies of flow in vessels of the forearm, the unmodified dimensions are particularly narrow but are not unusual \cite{Lee2012,Wilmink2018}. Furthermore, the geometries we are studying were obtained from a female, where narrower vessels again are not unexpected \cite{Corretti1995,Wilmink2018}. It can also be noted from \cite{Corretti1995} that the average flow rate in narrower vessels is also typically less than observed in larger vessels. The inlet velocity profile for our test cases (peak flow speed of 28.6 cm/s and average of 4.2 cm/s) is consistent with literature data e.g. \cite{Mitchell2004}, though this is a measure that can vary widely. The maximum velocity in the brachial artery is approximately 40 cm/s, achieved at a narrow throat shortly after the inlet plane which is also consistent with literature data \cite{Zambanini2005}. Given the narrow vessels present in our geometries, the brachial flow rate observed (9.8 ml/min) is naturally less then what may be reported in some cases within the literature but consistent with the lower bounds of these when stated standard deviations are taken into account \cite{Chambers2001,Corretti1995}. In the healthy geometry, the flow exiting the cephalic vein averaged 3.0 ml/min whilst again very low, this is not inconsistent with literature data once the geometry and flow characteristics of the test case is taken into account \cite{Albayrak2007}. \\

With the presence of a pre-maturation arteriovenous fistula, the flow at the exit plane of the cephalic vein increased by 50\% to 4.5 ml/min. In the post-maturation configuration, this flow rate has further increased to 5.6 ml/min, a 90\% increase compared to the healthy domain. Before attempting to compare this data with physical arteriovenous fistula, a number of points need to be taken into account. Firstly, these results were obtained with the arterial inlet velocity profile being identical to that used in the healthy case. This means that the increases in flow rates observed are solely due to the changes in vascular geometry caused by the creation of the fistula and the subsequent dilation of the cephalic vein. The 26\% increase in outflow due to the dilation of the cephalic vein may be indicative of this enlarged vessel drawing flow from the other veins in the domain. In our study, the arterial domain was not dilated in response to the maturation of the fistula and the connection point on the radial vein remained at the same diameter (the dilation be accounted for in the connecting bed model). This meant that approximately 31\% of the total arterial flow was passing through the fistula connection point, corresponding to 39\% of the flow passing along the radial artery. With the cross-sectional area of the AVF connection point being only 31\% of the radial outlet area, this would also suggest that the presence of the fistula enhances the flow exiting the arterial domain through this pathway. \\

\begin{figure}[!ht]
 \begin{subfigure}{\textwidth}
   \centering
   \includegraphics[width=0.75\textwidth]{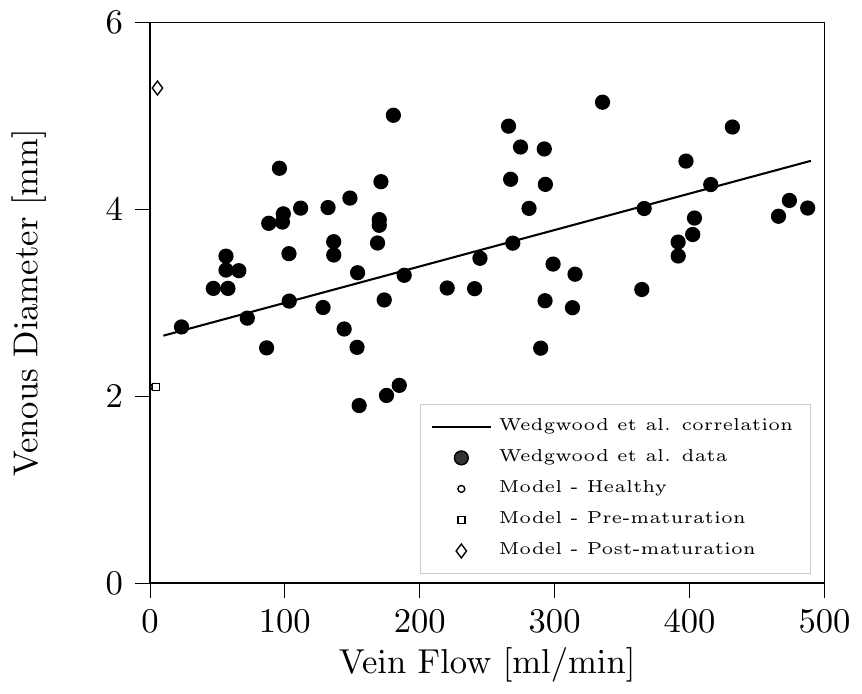}
   \caption{Comparison to full data from Wedgwood et al. \cite{Wedgwood1984}}
   \label{WedgewoodFull}
   \end{subfigure}
 \begin{subfigure}{\textwidth}
    \centering
   \includegraphics[width=0.75\textwidth]{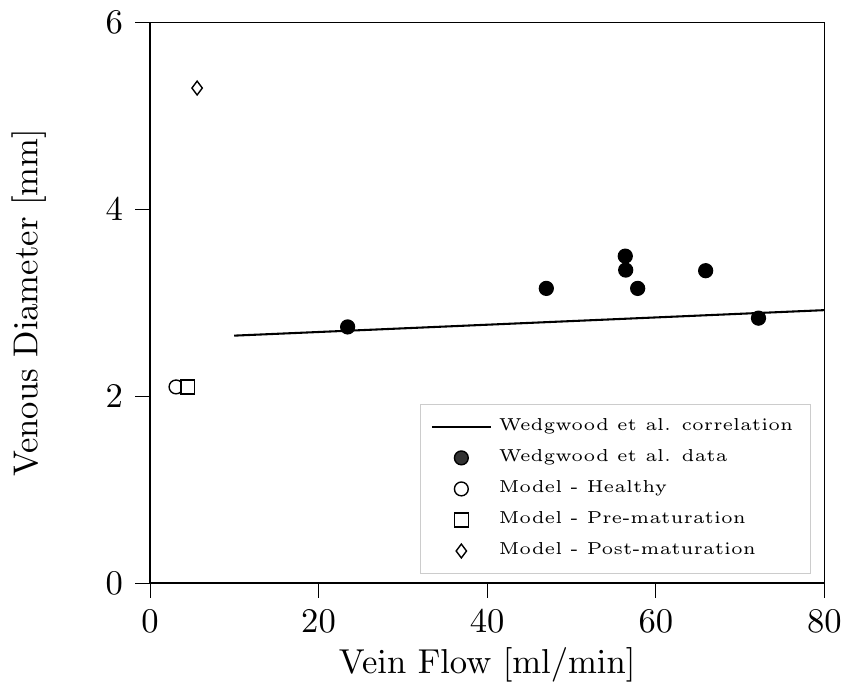}
   \caption{Comparison to data from Wedgwood et al. \cite{Wedgwood1984} below 80 ml/min}
   \label{WedgewoodZoom}
   \end{subfigure}
  \caption{Comparison of cephalic vein flow rate calculated by our model and a correlation generated in Wedgwood et al. \cite{Wedgwood1984}. In \subref{WedgewoodFull}, all data points of Wedgwood et al. have been indicated and the linear correlation  extended to encompass the both the observed data and that obtained from our simulations. In \subref{WedgewoodZoom} we zoom in on the low flow rate data in the vicinity of that observed in our simulation case.}
  \label{WedgwoodCorrelation}
\end{figure}

However, compared to observed data, the flow through the fistula recorded in our setting is far lower than what would be considered successful in a clinical setting. Such examples of `failed' fistula are not uncommon in practice, both in general and for patients with vessels of similar size to those tested here \cite{Wilmink2018}. In \cite{Wedgwood1984}, a study of 71 patients who had an arteriovenous fistula created between the radial artery and cephalic vein is presented. The authors present a correlation between the diameter of the vein used to create the fistula and the flow rate through it. As indicated in Figure \ref{WedgwoodCorrelation} the healthy and pre-maturation data is consistent with that the data presented in \cite{Wedgwood1984}, whilst the post-maturation flow rate is signficantly less than what would be expected for a vessel of this size. In this plot we have used the average of the inlet and  outlet radii of the cephalic vein. It is noted by Van Tricht et al. \cite{VanTricht2005} that, for a radiocephalic arteriovenous fistula, a flow rate of at least 500 ml/min is required to be successful (though \cite{Allon2002} notes that such a criteria varies globally). This is around 100 times greater than the cephalic flow observed due to the physiological input to the cases presented here and indicates how significantly the creation of a fistula alters the vasculature of the arm. To achieve such changes, flow feeding the brachial artery would need to similarly increase through a combination of increased mean velocity (such as through a change in velocity profile) or vascular dilation. As an example, changing the inlet profile to a steady maximum input velocity of 11.4 cm/s (corresponds to a mean of 5.5 cm/s) yielded an average flow rate of 12.7 ml/min entering the brachial vein and 6.1 ml/min exiting the cephalic vein in the pre-maturation configuration. Interestingly, the post-maturation cephalic flow was 5.9 ml/min using this input. This indicates that the pulsatility of flow also has an impact on how flow is distributed through the venous domain. Flow through the fistula could be additionally achieved by increased relative area of the connection point of the cephalic vein to the radial artery. As can be seen in \cite{Ogawa2011}, the brachial flow rate in patients with arteriovenous fistula can vary significantly between individual cases, making such behaviour difficult to predict in a modelling study without direct access to clinical data. \\

These factors point to some of the limitations of our current modelling approach. The LBM can be sensitive to the choice of discretisation parameters when it comes to the stability and accuracy of simulations and the numerical factors driving these can act in opposition to each other \cite{Kruger2017}, particularly in the vicinity of physiologically realistic flows. Significant increase in the flow velocity using the current configuration would lead to inaccuracies and instabilities in the LBM. Whilst these could, in principle, be improved through a reduction in the grid spacing, this would need to be balanced against the challenges of geometry generation, data storage and longer simulation wall times. This should motivate further efforts in HemeLB to incorporate more sophisticated collision kernels that offer options to simulate challenging flow regimes \cite{Kruger2017}. \\

In our study, we have only considered a single set of patient-specific vessels. Increasing the number of domains studied (both pre- and post-maturation) would serve to further streamline our procedure for \textit{in silico} fistula creation and allow us to develop further confidence in our approach for fistula simulation . \\ 

There are a number of further features of our current model that can be improved in future work. In particular, we plan to accurately integrate the behaviour of elastic walls into our model to better represent how they change in response to flow. We will also continue our work in extending the coupling approaches outlined here to enable circulatory flow to be captured within a full-human systemic geometry of arteries and veins that is ultimately coupled to a 3D model of the heart. The framework of the virtual human permits such increases in fidelity or performance to be readily introduced. The completion of these steps will constitute significant further progress along the pathway towards the virtual human. However, the culmination of this endeavour cannot be realised from a simulation perspective alone. The ongoing development of close collaborations between modellers and clinicians will be critical in ensuring that its significant benefits for predictive medicine, patient outcomes and general wellbeing become accessible to all members of society. \\

\section{Methods}
\label{sec:Methods}
We employ the LBM-based solver HemeLB \cite{HemeLBweb} to simulate macroscopic blood flow in coupled arterial and venous geometries. Further technical information on the LBM can be found in the many textbooks that cover this topic in detail (e.g.\ \cite{Succi2001,Mohamad2011,Guo2013,Kruger2017,Succi2018}).  HemeLB is an open-source code that has been optimised for the study of flow in sparse geometries such as those typical of vascular networks. Since its inception \cite{Mazzeo2008}, HemeLB has been used to study problems including flow validation in cerebral vessels \cite{Groen2018}, haemodynamic force distributions in retinal vessels \cite{Bernabeu2014} and wall shear stress analysis \cite{Bernabeu2012,Franco2015,Franco2016} in cerebral arteries, and the transport of magnetic nanoparticles for drug delivery within the circle of Willis \cite{Patronis2018}. These studies demonstrate how HemeLB, and 3D modelling techniques more generally, are able to accurately capture complex physiological flow phenomena in ways that are not possible with 1D approaches and can be difficult or impossible to measure and assess directly within a patient. In this work, we implement velocity inlet boundary conditions and pressure outlet conditions utilising the methods described in previous publications using HemeLB \cite{Nash2014}. We assume that solid walls are rigid with fluid-solid interfaces implemented using the technique of Bouzidi et al. \cite{Bouzidi2001}. \\

\begin{figure}[!ht]
   \centering
   \includegraphics[width=0.98\textwidth]{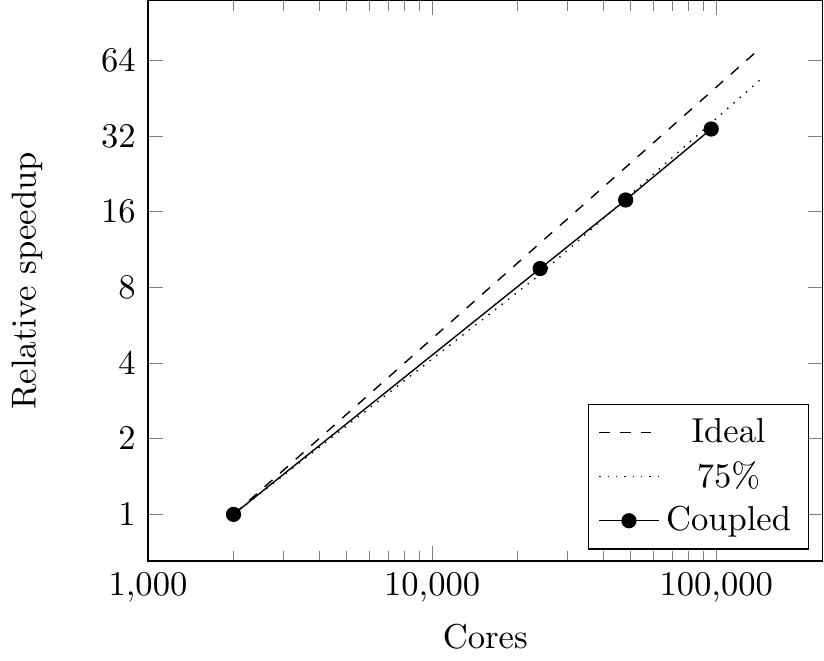} 
  \caption{Strong scaling performance of the self-coupled HemeLB implementation described in \cite{mccullough2020blood} to 96,000 cores of SuperMUC-NG. Ideal and 75\% scaling lines are provided as a guide. The coupled simulation transferred data between simulations every 500 steps.}
  \label{CoupledScaling}
\end{figure}

We build on the self-coupled HemeLB presented in our previous work \cite{mccullough2020blood}, which linked two instances of HemeLB in order to capture the respective flow in arterial and venous geometries. The scheme implemented here was designed to streamline the amount of data transferred between the two instances. In Figure \ref{CoupledScaling}, we demonstrated the strong scaling of the coupled model described in \cite{mccullough2020blood} as applied to a pair of circle of Willis geometries used in this work for the scaling of a single component version of HemeLB. In this case, we run the simulation for 5050 steps and perform coupling every 500 steps. This demonstrates that the coupled model exhibits strong scaling behaviour up to 96,000 cores on the SuperMUC-NG supercomputer SuperMUC-NG (at the Leibniz Supercomputing Centre, Germany).\\

\subsection{Sub-scale representation of capillary beds}
\label{sec:SubScale}
In this work we outline one method designed to represent the capillary beds linking the arterial and venous geometries that cannot be fully resolved for macroscopic simulations. This begins by first mapping the arterial outlets to the venous inlets. The separate collections of these represent individual beds. For any given bed, the sub-scale model needs to be able to accommodate an arbitrary number of arteries and veins being connected. In most regions of the body, there are significantly more veins than arteries of simulatable scale. This means that most beds will have $n$ arteries and $m$ veins connected where typically $n<m$. Local vascular features, however, may result in alternative configurations. In the current work we identify bed regions and use the sub-scale model to represent the capillaries connecting the arteries and veins located within these regions. Within this model, we use the average velocities from vessels on the opposite geometry to update those on the current set of vessels. The change in velocity here represents the action of the sub-scale capillary network on the flow. Coupling occurs at specified simulation intervals and we alternate the direction of coupling at each instance. As the LBM represents a quasi-incompressible fluid, we use the updated velocities on the arterial geometry to apply a force to the outlets based on the dynamic pressure at that outlet. This represents the volume of fluid within the sub-scale region. This forcing term is applied at the outlet using the approach of Guo et al. \cite{Guo2002}. \\

The capillary bed approach used in this paper  is based on resistor-capacitor models used in some existing models as capillary beds or windkessel boundary conditions. We have taken particular inspiration from the works of Mynard et al. \cite{Mynard2015a,Mynard2015b}. In our approach we firstly determine the flow weighted velocity, $v_D$ on the arterial and venous sides of a capillary bed 

\begin{equation}
v_D = \sum_j\frac{A_jv_j^2}{A_jv_j}.
\end{equation}

\begin{figure}
\includegraphics[width=0.98\textwidth]{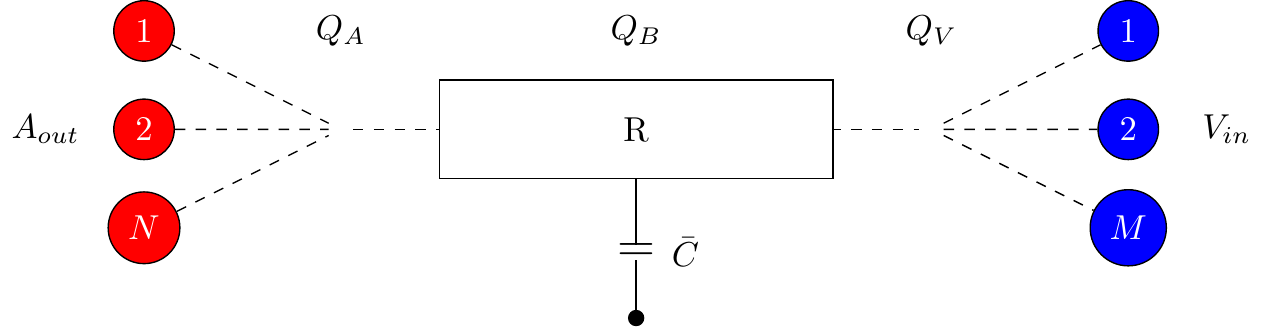}
\caption{Conceptual layout of the resistor-capacitor based approach to representing the capillary beds. The arterial, $Q_A$ and venous, $Q_V$ flow through the bed is accumulated from and distributed to the connecting vessels by their relative portion of the total vessel area attached to the bed. The capacitance, $\bar{C}$, can be viewed as an average of arterial and venous vessel capacitances.}
\label{RCconcept}
\end{figure}

\noindent The flow rate, $Q$, can then be determined by scaling $v_D$, by the total area of the appropriate vessels. We then consider the capillary bed as the conceptual model presented in Figure \ref{RCconcept}. Here the bed has capacitance, $\bar{C}$, and resistance, $R$. The change in pressure over the bed is computed as $\Delta P = RQ_B$ where $Q_B$ is the flow through the bed itself. Taking capacitance into account, the flow of the system can be described as

\begin{equation}
Q_A - \bar{C}\frac{d\bar{P}}{dt} = Q_V,
\end{equation}

\noindent where $\bar{P} = 0.5(P_A + P_V)$ is the average pressure of the system. We note that this can be recast as $\bar{P} = P_A - 0.5(P_A - P_V) = P_A - 0.5RQ_B$. We assume that the majority of the pressure change arises on the venous side of the system, leading to

\begin{equation}
\frac{d\bar{P}}{dt} \approx \frac{d}{dt} -0.5RQ_B = -0.5R\frac{dQ_B}{dt}.
\end{equation}

\noindent This then results in
\begin{equation}
Q_A + 0.5 \bar{C}R\frac{dQ_B}{dt} = Q_V,
\end{equation}

\noindent which can be used to update the flow rate on the receiving side of the capillary coupling. This is then distributed to the individual vessels based on the relative area of each. The implementation of this method in HemeLB can be obtained from \url{https://github.com/UCL-CCS/HemePure_SelfCoupled/tree/ResistorCapacitorBeds}. \\

\bibliographystyle{unsrtnat_JMmod}
\bibliography{2020_CouplingModelsPaper}

\section*{Funding}
We acknowledge funding support from European Commission CompBioMed Centre of Excellence (Grant No. 675451 and 823712). Support from the UK Engineering and Physical Sciences Research Council under the project `UK Consortium on Mesoscale Engineering Sciences (UKCOMES)' (Grant No. EP/R029598/1) is gratefully acknowledged. We acknowledge funding support from MRC for a Medical Bioinformatics grant (MR/L016311/1), and special funding from the UCL Provost. \\

The authors gratefully acknowledge the Gauss Centre for Supercomputing e.V. (\url{www.gauss-centre.eu}) for funding this project by providing computing time on the GCS Supercomputer SuperMUC-NG at Leibniz Supercomputing Centre (\url{www.lrz.de}). \\

\section*{Acknowledgements}
The authors acknowledge the Foundation for Research on Information Technologies in Society (IT'IS), particularly B. Lloyd and E. Neufeld, for providing the anatomical geometries used in this project as part of a collaboration within the CompBioMed Centre of Excellence.

\section*{Author Contributions}
JM developed the coupling model used here, devised, ran and analysed the simulations and drafted the initial version of the manuscript. PC supervised the project, provided scientific and technical advice and contributed to writing the manuscript.

\section*{Additional Information}
The authors declare no competing interests.

\listoffigures

\end{document}